\documentclass[american,aps,pra,reprint,nofootinbib]{revtex4-2}
\usepackage[T1]{fontenc}
\usepackage{amsmath,amssymb,amsthm,mathtools,booktabs,array}
\usepackage[hidelinks]{hyperref}
\usepackage{xcolor}

\newtheorem{theorem}{Theorem}
\newtheorem{lemma}[theorem]{Lemma}
\newtheorem{proposition}[theorem]{Proposition}
\newtheorem{corollary}[theorem]{Corollary}
\theoremstyle{definition}
\newtheorem{definition}[theorem]{Definition}
\theoremstyle{remark}

\newcommand{\Tr}{\operatorname{Tr}}

\newcommand{\SEP}{\operatorname{SEP}}

\newcommand{\Emax}{E_{\max}}
\newcommand{\ED}{E_D}
\newcommand{\cc}{\mathrm{cc}}

\newcommand{\Ph}{\Phi_2}

\newcommand{\Fcc}{F^{\cc}}

\usepackage{xcolor}
\definecolor{linkpink}{HTML}{EC008C}

\usepackage{hyperref}
\hypersetup{
    colorlinks=true,
    linkcolor=linkpink,  
    citecolor=linkpink,  
    urlcolor=linkpink    
}

\begin{document}
\title{Correlated Catalysis Removes the Strong Converse for Entanglement Distillation}

\author{Piotr Masajada}
\author{Aby Philip}
\author{Alexander Streltsov}
\email{streltsov.physics@gmail.com}
\affiliation{Institute of Fundamental Technological Research, Polish Academy of Sciences, Pawi\'{n}skiego 5B, 02-106 Warsaw, Poland}

\begin{abstract}
Entanglement distillation converts noisy shared quantum states
into maximally entangled pairs using local operations and
classical communication. The distillable entanglement $E_D$ determines
the maximum asymptotic rate at which this process can be
performed reliably. For several classes of quantum states,
distillation above this rate obeys a strong converse: the
fidelity with the target state vanishes asymptotically.
Here we show that correlated catalysis removes the
strong converse for every distillable bipartite state.
Specifically, at any target rate $R>E_D$, the optimal asymptotic
fidelity is given by
$E_D/R$.
Thus, although catalysis does not increase the maximum
reliable distillation rate, it allows nonvanishing fidelity
at every finite rate. Remarkably, this advantage persists
even when the catalyst is returned exactly at every finite
blocklength and the correlations between the output and
catalyst vanish asymptotically. We further characterize the catalyst size required
to sustain nonvanishing fidelity above the reliable distillation
rate. 
\end{abstract}

\maketitle

\section{Introduction}
Quantum entanglement~\cite{Horodeckireview} is a fundamental type of correlation between quantum particles, enabling communication protocols that surpass classical limits. As a uniquely quantum phenomenon, entanglement acts as a vital resource for tasks such as quantum teleportation~\cite{Benettteleportofstate}, superdense coding~\cite{Benettsuperdensecoding}, and quantum key distribution~\cite{BennettQKD,Ekertqkd}. These applications motivate the development of methods for preparing high-quality entangled states.

Entanglement is often studied in the context of resource theory. In this setting, two parties, say Alice and Bob, can perform only local quantum channels, and communicate through a classical channel. These operations are called local operations and classical communication (LOCC)~\cite{Chitambarlocc}. 

Entanglement distillation enables the extraction of nearly maximally entangled states from noisy entangled input using local operations and classical communication~\cite{BennettBBPSSW,Bennetterrordistillation,DeutschDEJMPS,Dehaenedistillation,Horodeckihigherdistillation,Dürdistillation,Matsumotodistillation}. For a finite number of input copies, both deterministic protocols based on quantum error correction~\cite{Bennetterrordistillation,Matsumotodistillation} and probabilistic protocols based on postselection~\cite{BennettBBPSSW,DeutschDEJMPS} have been studied. In the asymptotic regime, the central objective is to determine the optimal rate at which maximally entangled pairs can be extracted as the number of input copies tends to infinity~\cite{Bennetpuredistillation,Bennetterrordistillation,Devetak2005}.

In the asymptotic regime, a distillation protocol $\mathcal{D}_{n}$ transforms \(n\) copies of a bipartite state \(\rho\) into an output state approximating \(m_n\) maximally entangled qubit pairs. The distillation rate is the number of target pairs per input copy, \(R=\lim_{n\to\infty}m_n/n\). The distillable entanglement \(E_D(\rho)\) is the supremum of rates achievable such that the fidelity between the output state and $\lfloor Rn\rfloor$ maximally entangled states tends to unity~\cite{Rainsdistillable}: 
\begin{equation}
\ED(\rho)=\sup\left\{
R\geq 0:
\limsup_{n\to\infty}
\Tr\left[
\Ph^{\otimes\lfloor Rn\rfloor}
\mathcal{D}_n(\rho^{\otimes n})
\right]=1
\right\},
\label{eq:distillable_entanglement}
\end{equation}
where the supremum is taken over rates \(R\) for which there exists a sequence of LOCC channels \(\{\mathcal{D}_n\}_{n\in\mathbb{N}}\), and \(\Ph\) denotes the density operator of a maximally entangled qubit pair. Operationally, \(E_D(\rho)\) quantifies the asymptotic number of Bell pairs that can be extracted per input copy with vanishing error~\cite{Pleniomeasures}.  

The definition of distillable entanglement rules out asymptotically perfect distillation at rates \(R>E_D(\rho)\), but does not determine how the fidelity behaves in this regime. A stronger requirement is the strong converse property:
for every rate $R>E_D(\rho)$, the fidelity of the output with
the target state $\Ph^{\otimes\lfloor Rn\rfloor}$ must
tend to zero as $n\to\infty$, irrespective of the sequence
of LOCC protocols employed~\cite{Lamisecondlaw}. Thus, when the strong converse holds, allowing a fixed nonzero error does not increase the asymptotically achievable distillation rate. 

It is known that the strong converse holds for bipartite pure states~\cite{Masahitoconverse} and for maximally correlated states~\cite{Fangconverse}. More generally, the Rains relative entropy provides a strong-converse upper bound on the distillation rate: whenever the rate exceeds this bound, the fidelity vanishes exponentially with the number of input copies \cite{Tomamichelconverse,Khatripqct}. Whether the strong converse holds for unassisted distillation from arbitrary bipartite states remains open. In this paper, we investigate whether the strong converse also holds for entanglement distillation protocols assisted by a catalyst.

Entanglement catalysis provides another way to extend the possibilities of quantum state manipulation under LOCC~\cite{Jonathancatalysis,Dattacatalysis,Lipkacatalysis,Eisertcatalysis,Nielsenquantumcomputation}. In its original formulation~\cite{Jonathancatalysis}, Alice and Bob have access to an auxiliary entangled state \(\omega_C\), called a catalyst, which assists a transformation while being returned unchanged and uncorrelated with the output. A catalytic transformation therefore takes the form 
\begin{equation}
    \Lambda_{\mathrm{LOCC}}(\rho\otimes\omega_C)=
    \sigma\otimes\omega_C.
\end{equation}
Remarkably, such a transformation may be possible even when \(\rho\) cannot be converted into \(\sigma\) by LOCC alone. This framework was subsequently extended to correlated catalysis~\cite{Wilmingcatalysis,Shiraishicorrelated,Wilmingcorrelated}, in which the catalyst is returned exactly in its marginal state but may become correlated with the output system~\cite{Kondracatalysis,Ganardicatalysis,Lamicorrelated}. The final state \(\Omega_{SC}\) is such that 
\begin{equation}
\Omega_{SC}=\Lambda_{\mathrm{LOCC}}(\rho\otimes\omega_C),
\qquad
\Omega_S=\sigma,
\qquad
\Omega_C=\omega_C,
\end{equation}
 instead of \(\Omega_{SC}=\sigma\otimes\omega_C\). Allowing such residual correlations can enlarge the set of achievable state transformations~\cite{Kondracatalysis}.

Entangled states
with positive partial transpose remain undistillable
under correlated catalytic local operations and classical
communication, while retaining a positive entanglement
cost~\cite{Lamicorrelated}.
Thus, correlated catalysis does not make entanglement
theory asymptotically reversible.

In this work, we show that correlated catalysis removes the strong converse for every distillable bipartite state, even when the correlations between the output and the catalyst vanish asymptotically. More precisely, at any rate \(R>E_D(\rho)\), the optimal asymptotic fidelity equals \(E_D(\rho)/R\), whereas fidelity approaching unity is achievable for \(R\leq E_D(\rho)\). These limits remain achievable under any prescribed positive sequence of upper bounds on the output–catalyst mutual information tending to zero.

We also determine the catalytic resources needed to sustain
non-vanishing fidelity above the reliable distillation rate.
At rates where non-catalytic distillation obeys an exponential
strong converse, the catalyst's max-relative entropy of
entanglement must grow at least linearly with the input
blocklength $n$.
Requiring the output–catalyst mutual information to satisfy \(I(S:C)\leq\delta_n\), for any prescribed positive sequence \(\delta_n\to0\), strengthens this lower bound to $\Omega(n\log_2(1/\delta_n))$, and hence imposes the same lower bound on the logarithm of each local catalyst dimension.
Conversely, for every distillable source, every
$R>E_D(\rho)$, and every fixed $0<f_0<E_D(\rho)/R$,
we construct correlated catalytic protocols achieving
fidelity at least $f_0$ with
$\log_2\dim C_A,\log_2\dim C_B=O(n\log_2(1/\delta_n))$.
Thus, whenever the exponential-converse assumption holds,
the optimal catalyst dimension satisfies
$\log_2 D_n^{\mathrm{opt}}=\Theta(n\log_2(1/\delta_n))$.
In particular, this conclusion applies to pure and maximally
correlated source states at every rate above their distillable
entanglement.

\section{Definitions and problem statement}
\noindent Let 
\begin{equation}
    |\Phi_d\rangle=\frac{1}{\sqrt d}\sum_{x=0}^{d-1}|x\rangle_A|x\rangle_B,
 \qquad \Phi_d=|\Phi_d\rangle\langle\Phi_d|
\end{equation}
be a maximally entangled state with local dimensions $d$. The $m$-fold tensor product of a maximally entangled qubit pair is therefore $\Ph^{\otimes m}=\Phi_{2^m}$. 

As discussed in the introduction, catalysis can enable transformations that are otherwise impossible under LOCC alone. Here, we study entanglement distillation assisted by a bipartite catalyst. We define a distillation protocol with catalysis as follows~\cite{Kondracatalysis,Ganardicatalysis}:
\begin{definition}{\label{def::protocol}}[Correlated catalytic distillation protocol]
Fix a bipartite source state $\rho$ and a target of $m$ Bell pairs. A correlated catalytic distillation protocol is a deterministic LOCC channel $\mathcal{D}_{dist}$ that takes $n$ copies of the state $\rho$ and a bipartite catalyst $\omega_C$ and outputs
\begin{equation}
 \Omega_{SC}=\mathcal{D}_{dist}\!\left(\rho^{\otimes n}\otimes\omega_C\right),
 \qquad \Omega_C=\omega_C.
 \label{eq:exactreturn}
\end{equation}
We quantify the performance of a catalytic distillation protocol by the overlap of its output marginal with the target maximally entangled state:
\begin{equation}
 f(\Omega_{SC},m)\coloneqq\Tr\!\left[(\Phi_{2^m}\otimes I_C)\Omega_{SC}\right]
 =\Tr(\Phi_{2^m}\Omega_S).
 \label{eq:globalfidelity}
\end{equation}  
\end{definition}

The correlations between the output and the catalyst can be quantified by the quantum mutual information~\cite{Wildeqit}
\begin{equation}
    I(S:C)_\Omega
=S(\Omega_S)+S(\omega_C)-S(\Omega_{SC}),
\end{equation}
where \(S(\rho)=-\mathrm{Tr}(\rho\log(\rho))\) denotes the von Neumann entropy~\cite{Nielsenquantumcomputation}. Additionally, we can bound the trace distance to uncorrelated catalysis by the quantum Pinsker inequality with quantum mutual information~\cite{Hiaiinequality},
\begin{equation}
 \frac12\|\Omega_{SC}-\Omega_S\otimes\omega_C\|_1
 \le\sqrt{\frac{\ln2}{2}I(S:C)_\Omega}.
 \label{eq:pinsker}
\end{equation}
In particular, $I(S:C)_\Omega=0$ implies
$\Omega_{SC}=\Omega_S\otimes\omega_C$.
Moreover, vanishing mutual information implies
that the joint state approaches the product of
its marginals in trace distance.

Throughout this work, we consider finite-dimensional catalysts with unrestricted dimension and entanglement, and investigate protocols in which the residual correlations between the system $\Omega_S$ and catalyst $\Omega_C$ can be made arbitrarily small while preserving the exact catalyst marginal.

Suppose that we have $n$ copies of a state $\rho$ and want to produce $m$ copies of $\Ph$ using correlated catalytic entanglement distillation.
The maximum achievable fidelity with a target state $\Phi_{2^{m}}$ is:
\begin{equation}
\Fcc_{\rho\to\Ph}(n,m;\delta)
=
\sup_{\omega_C,\mathcal{D}_{dist}}
\Tr\left(\Ph^{\otimes m}\Omega_S\right),
\label{eq:optimal_fidelity}
\end{equation}
where the supremum is taken over all finite-dimensional
catalysts and deterministic LOCC channels satisfying
$\Omega_C=\omega_C$ and $I(S:C)_\Omega\leq\delta$. See Eq. \eqref{eq:exactreturn} for the definition of $\Omega_{SC}.$
We use $\delta=\infty$ when no restriction on the
output--catalyst correlations is imposed.

For a fixed distillation rate $R>0$, let
$m_n=\lfloor Rn\rfloor$. The reliable catalytic
distillation rate is defined as
\begin{equation}
D^{\cc}(\rho)
=
\sup\left\{
R:
\limsup_{n\to\infty}
\Fcc_{\rho\to\Ph}(n,m_n;\infty)=1
\right\}.
\label{eq:reliable_rate}
\end{equation}
This quantity coincides with the correlated-catalytic
distillable entanglement
$E_{d,\mathrm{LOCC}^{cc}}(\rho)$ defined
in~\cite{Lamicorrelated}, since vanishing trace-distance
error and fidelity tending to one are equivalent
for the pure target state.
The corresponding strong-converse threshold is
\begin{equation}
\widetilde{D}^{\cc}(\rho)
=
\sup\left\{
R:
\limsup_{n\to\infty}
\Fcc_{\rho\to\Ph}(n,m_n;\infty)>0
\right\}.
\label{eq:strong_converse_threshold}
\end{equation}

The strong converse holds if every rate exceeding the
reliable distillation rate necessarily leads to
vanishing global fidelity in the limit $n\rightarrow \infty.$ In other words, for an input state $\rho,$ the strong converse holds iff $D^{cc}(\rho)=\widetilde {D}^{cc}(\rho).$  In this paper we show that strong converse does not hold for catalytic entanglement distillation protocols. 

Throughout this work, all logarithms are base two
unless stated otherwise. We use standard asymptotic
notation: for nonnegative sequences $a_n$ and $b_n$,
we write $a_n=O(b_n)$ if there exists a constant
$C>0$ such that $a_n\le Cb_n$ for all sufficiently
large $n$.
We write $a_n=\Omega(b_n)$ if there exists a constant
$c>0$ such that $a_n\ge cb_n$ for all sufficiently
large $n$.
Finally, we write $a_n=\Theta(b_n)$ if there exist
constants $c,C>0$ such that
$cb_n\le a_n\le Cb_n$ for all sufficiently large $n$.

Throughout this work, fidelity refers to the squared fidelity;
for a pure target $\Phi$, it is given by $\Tr(\Phi\rho)$.

\section{Main results}

We first show that exact catalysis with vanishing correlations removes
the strong converse for every distillable source state.
At any rate $R>E_D(\rho)$, the optimal asymptotic 
fidelity is $E_D(\rho)/R$, even when output--catalyst
correlations vanish asymptotically.

\begin{theorem}
\label{thm:mixedmain}
Let $\rho$ be a finite-dimensional bipartite state with
$\ED(\rho)>0$.  For every prescribed positive sequence
$\delta_n\to0$ and every $R>0$,
\begin{equation}
 \lim_{n\to\infty}
 \Fcc_{\rho\to\Ph}\!\left(n,\lfloor Rn\rfloor;\delta_n\right)
 =\min\left\{1,\frac {\ED (\rho)} {R}\right\}.
 \label{eq:mixedcurve}
\end{equation}
Consequently,
\begin{equation}
 \widetilde D^{\cc}(\rho)=\infty.
 \label{eq:mixedrates}
\end{equation}
\end{theorem}
\noindent The proof is given in Section~\ref{se::mixedmain}.
In particular, if $\rho=\Phi_2$ we obtain 
\begin{align}
\lim_{n\to\infty}\Fcc_{\Ph\to\Ph}\!\left(n,\lfloor Rn\rfloor;\delta_{n}\right) & =\min\left\{ 1,\frac{1}{R}\right\}, \label{eq:bellcurve}\\
\widetilde{D}^{\cc}(\Ph) & =\infty.\label{eq:bellrates}
\end{align}
Note that the result
$D_{\mathrm{cc}}(\rho)=E_D(\rho)$ was previously established
for distillable states in Ref.~\cite{Ganardicatalysis}.

The construction used in the proof of Theorem~\ref{thm:mixedmain} achieves
the optimal asymptotic fidelity $E_D(\rho)/R$ at rates
$R>E_D(\rho)$, but requires a large catalyst memory.
In the following, we derive lower bounds on the catalyst resources
needed to maintain a fixed positive fidelity at rates
where non-catalytic distillation obeys an exponential
decay of fidelity.

Let 
\begin{equation}
    F_{\rho\to \Ph}^{\text{not}}(n,m)=\sup_{\Lambda\in LOCC}\mathrm{Tr}\left( \Phi_{2^{m}} \Lambda(\rho^{\otimes n}) \right)
\end{equation}
denote the optimal fidelity without a catalyst. 
For a bipartite catalyst state $\omega_C$, define its max-relative entropy of entanglement~\cite{Datteentropies}
\begin{equation}
\begin{split}
    &\Emax(\omega_C)=\inf_{\tau\in\SEP}D_{\max}(\omega_C\Vert\tau),\\
 &D_{\max}(x\Vert y)=\inf\{\lambda:x\le2^\lambda y\}.
\end{split}
 \label{eq:Emaxdef}
\end{equation}
The Rains relative entropy
$\mathcal{R}_{ent}(\rho),$ defined as~\cite{Rainsentropy}
\begin{equation}
    \mathcal{R}_{ent}(\rho)
    :=
    \inf_{\substack{\sigma\geq 0\\
                    \|\sigma^{T_B}\|_1\leq 1}}
    D(\rho\|\sigma),
    \label{eq::rainsentropy}
\end{equation}
 provides an exponential converse-bound
 for non-catalytic distillation of any
finite-dimensional bipartite state~\cite{Tomamichelconverse,Khatripqct}, i.e. every rate $R$ satisfying $R>\mathcal{R}_{ent}(\rho)$ leads to exponential decay of $F_{\rho\to\Ph}^{\text{not}}.$  In Eq. \eqref{eq::rainsentropy}, $T_B$ denotes partial transposition on Bob's
subsystem~\cite{Horodeckipptcriterion,Perespptcriterion} and
$D(\rho\|\sigma)=\operatorname{Tr}[
\rho(\log_2\rho-\log_2\sigma)]$ is the quantum
relative entropy, with value $+\infty$ when
$\operatorname{supp}\rho\not\subseteq
\operatorname{supp}\sigma$.

For pure and maximally correlated states,
$E_D(\rho)=\mathcal{R}_{ent}(\rho)$~\cite{Fangconverse}, so for these states every rate above distillable entanglement leads to exponential decay of fidelity with target state $\Ph$. Our catalyst-resource lower bounds require an exponential fidelity bound at a threshold strictly below the target rate. We formulate this assumption below. 

\begin{definition}[Exponential Converse Bound]{\label{de::conversebound}}
    Let \(\rho\) be a finite-dimensional bipartite state. We say that non-catalytic distillation from \(\rho\) satisfies an exponential-converse bound with parameters \(r_{\mathrm{sc}}\geq0\), \(\gamma>0\), and \(N_0\in\mathbb N\) if
\begin{equation}
    F^{\mathrm{not}}_{\rho\to\Phi_2}(N,L)\leq 2^{-\gamma N}
    \label{eq::strongexpconverse}
\end{equation}
for every integer \(N\geq N_0\) and every integer \(L\geq r_{\mathrm{sc}}N\). The parameters are independent of \(N\) and \(L\).
\end{definition}

We next derive lower bounds on the catalyst entanglement required to maintain nonvanishing fidelity at rates where non-catalytic distillation obeys an exponential converse bound. The following theorem applies at every target rate $R>\mathcal{R}_{\mathrm{ent}}(\rho)$, since its exponential-converse assumption is satisfied by choosing a threshold rate $r_{\mathrm{sc}}$ such that $\mathcal{R}_{\mathrm{ent}}(\rho)<r_{\mathrm{sc}}<R$.

\begin{theorem}[Necessary catalyst entanglement]
\label{thm::exp}
 Let \(\rho\) be a finite-dimensional bipartite state satisfying the exponential-converse bound of Definition \ref{de::conversebound} with parameters \(r_{\mathrm{sc}}\), \(\gamma\), and \(N_0\). Consider a deterministic catalytic LOCC protocol with $\rho^{\otimes n}$ as input, $\omega_{C}$ as catalyst such that it achieves the rate $R>r_{\text{sc}}$ and the fidelity of this protocol with a target Bell state is greater than a constant $f_0\in(0,1)$.

Then, for all sufficiently large $n$ and with no restriction on catalyst correlation, the catalyst’s max-relative entropy of entanglement satisfies
\begin{equation}
\label{eq:emax-linear}
    E_{\max}(\omega_C)
    \ge
    \gamma n-\log_2(1/f_0).
\end{equation}

Furthermore, if the correlations of the joint output state $\Omega_{SC}$  are restricted by
\begin{equation}
    I(S:C)_{\Omega}\le\delta_n,
    \qquad
    0<\delta_n\longrightarrow0,
\end{equation}
then
\begin{equation}
\label{eq:emax-correlation-cost}
    E_{\max}(\omega_C)
    =
    \Omega\!\left(
        n\log_2\frac1{\delta_n}
    \right).
\end{equation}
\end{theorem}

\noindent The proof is given in Section~\ref{se::exp}.

We next translate the lower bounds on \(E_{\max}(\omega_C)\) into lower bounds on the local catalyst dimensions.
Let
\begin{equation}
    D_n := \min\{\dim C_A,\dim C_B\}
\end{equation}
denote the smaller local catalyst dimension. We show the following corollary: 
\begin{corollary}[Necessary catalyst dimension]
\label{cor:catalyst-dimension}
Under the assumptions of
Theorem~\ref{thm::exp} and without a correlation constraint, the following lower bound on the catalyst dimension holds:
\begin{equation}
    D_n\ge f_0\,2^{\gamma n}
\end{equation}
for all sufficiently large $n$.

Moreover, if we also assume the correlation constraint 
$I(S_n:C_n)\le\delta_n\to0$, we obtain
\begin{equation}
    \log D_n
    =
    \Omega\!\left(
        n\log\frac1{\delta_n}
    \right).
    \label{eq::dimbound}
\end{equation}
Equivalently, there exists a constant $c>0$ such that
\begin{equation}
    D_n
    \ge
    \left(\frac1{\delta_n}\right)^{cn}
\end{equation}
for all sufficiently large $n$.
\end{corollary}

\begin{proof}

The max-relative entropy of entanglement satisfies
the standard dimension bound~\cite{Datteentropies,Dattaalias}
\begin{equation}
    E_{\max}(\omega_C)
    \le
    \log_2\min\{\dim C_A,\dim C_B\}.
\end{equation}
Consequently,
\begin{equation}
    \log_2 D_n\ge E_{\max}(\omega_C).
\end{equation}
Combining this inequality with Theorem~\ref{thm::exp} gives
\begin{equation}
    D_n
    \ge
    2^{\gamma n-\log_2(1/f_0)}
    =
    f_0\,2^{\gamma n}
\end{equation}
for protocols without a constraint on output–catalyst correlations.
Under the additional correlation
assumptions, Theorem~\ref{thm::exp} also yields
\begin{equation}
    \log_2 D_n
    =
    \Omega\!\left(n\log_2\frac1{\delta_n}\right),
\end{equation}
which proves the remaining claims.
\end{proof}

Although the above corollary provides a lower bound on the dimension of the catalyst, it does not provide a protocol that uses such a catalyst. The following theorem constructs a protocol which reaches the scaling specified by the lower bound on catalyst dimensions \eqref{eq::dimbound} of Corollary~\ref{cor:catalyst-dimension}. 
\begin{theorem}[Achievable catalyst dimension]
\label{thm::achieveexp}
Let $\rho$ be a finite-dimensional bipartite state
with $E_D(\rho)>0$. Fix
\begin{equation}
    R>E_D(\rho),
    \qquad
    0<f_0<\frac{E_D(\rho)}{R},
\end{equation}
and let $\delta_n>0$ be any prescribed sequence
with $\delta_n\to0$.

Then, for all sufficiently large $n$, there exists a
deterministic correlated catalytic LOCC protocol
with input $\rho^{\otimes n}$ and target
$\Phi_{2^{m_n}}$, where $m_n=\lfloor Rn\rfloor$,
such that
\begin{equation}
    \Omega_C=\omega_C,
    \,
    f_n:=\operatorname{Tr}
    \bigl(\Phi_{2^{m_n}}\Omega_S\bigr)\ge f_0,
    \,\,
    I(S:C)_\Omega\le\delta_n.
\end{equation}
Moreover, both local catalyst dimensions satisfy
\begin{equation}
    \log_2\dim C_A,\ \log_2\dim C_B
    =
    O\!\left(n\log_2\frac1{\delta_n}\right).
\end{equation}
\end{theorem}

\noindent The proof is given in Section~\ref{se::achieveexp}.

Now, let $D_n^{\mathrm{opt}}$ be the minimum value of
$\min\{\dim C_A,\dim C_B\}$ over all deterministic
correlated catalytic LOCC protocols with input
$\rho^{\otimes n}$, target $\Phi_{2^{\lfloor Rn\rfloor}}$,
fidelity at least $f_0$, and output--catalyst mutual
information at most $\delta_n$.
Using above results we show the following corollary.
\begin{corollary}[Optimal order of catalyst dimension]
\label{cor:optimal-catalyst-dimension}
Under the assumptions of
Theorem~\ref{thm::achieveexp},
suppose additionally that the exponential-converse
hypothesis of Theorem~\ref{thm::exp} holds for some
threshold strictly below $R$. Assume that the output-catalyst mutual information is at most $\delta_n.$ 
Then
\begin{equation}
    \log_2 D_n^{\mathrm{opt}}
    =
    \Theta\!\left(n\log_2\frac1{\delta_n}\right).
\end{equation}
\end{corollary}

\begin{proof}
Corollary~\ref{cor:catalyst-dimension} gives
\begin{equation}
    \log_2 D_n^{\mathrm{opt}}
    =
    \Omega\!\left(n\log_2\frac1{\delta_n}\right).
\end{equation}
The matching upper bound follows from
Theorem~\ref{thm::achieveexp}.
\end{proof}

The constants in these bounds may depend on $\rho$, $R$,
$f_0$, and the fixed exponential-converse parameters,
but not on $n$ or $\delta_n$.
The strict inequality $f_0<E_D(\rho)/R$ provides the
fixed positive gap required by our construction.
Thus, under the assumptions of Corollary~\ref{cor:optimal-catalyst-dimension}, the optimal
catalyst dimension satisfies
$\log_2 D_n^{\mathrm{opt}}
=\Theta(n\log_2(1/\delta_n))$
for every fixed fidelity requirement strictly below
the optimal asymptotic fidelity.
This does not determine the optimal dimension scaling
when the fidelity itself converges to $E_D(\rho)/R$.

Operationally, this corresponds to an optimal local
catalyst memory cost of $\Theta(n\log_2(1/\delta_n))$
qubits: every admissible protocol requires
$\Omega(n\log_2(1/\delta_n))$ qubits on each side,
and our construction achieves the matching upper bound
on both sides.
This counts the memory needed to represent the catalyst,
not the number of Bell pairs it contains.

\section{Proof of Theorem \ref{thm:mixedmain}}{\label{se::mixedmain}}
In this section, we detail the proof of theorem~\ref{thm:mixedmain}. Within the proof, we will use several lemmas whose full statements and proofs can be found in Section \ref{se::theorem2lemmas}.

To begin, assume that we have a state $\rho$ such that $\ED (\rho)>0$. First, we find an upper bound on $\limsup_{n\to\infty}
 \Fcc_{\rho\to\Ph}(n,m_n;\infty).$

Let a catalytic protocol take $\rho^{\otimes n}$ as input and produce an $m_n$-pair output of fidelity $f(\Omega_{SC},m_n)$.  Lemmas~\ref{lem:EDoutput} and \ref{lem:filter} give, for $m_n\ge8$,
\begin{multline}
    \bigl(m_n-\lceil2\log m_n\rceil\bigr)f(\Omega_{SC},m_n)-2
 \\\le \ED(\Omega_S)\le nE_D(\rho).
\end{multline}
Hence
\begin{equation}
 f(\Omega_{SC},m_n)\le
 \frac{nE_D(\rho)+2}{m_n-\lceil2\log m_n\rceil}.
 \label{eq:mixedfiniteupper}
\end{equation}
For $m_n=\lfloor Rn\rfloor$ this yields 
\begin{equation}
 \limsup_{n\to\infty}
 \Fcc_{\rho\to\Ph}(n,m_n;\infty)
 \le\min\left\{1,\frac{ E_D(\rho)}{R}\right\}.
 \label{eq:mixedupper}
\end{equation}
The bound is independent of catalyst dimension, catalyst entanglement, and output--catalyst correlation.

We now construct a protocol that asymptotically attains this upper bound. The protocol consists of two stages.
First, correlated catalytic amplification
produces a mixture of $\widetilde m>n$ copies
of the source state and a separable state.
Second, ordinary LOCC distillation with a rate $r$ is applied
to obtain $m_n$ Bell pairs. In the following paragraph we describe the first stage.

Fix a rate $r$ with $0<r<\ED(\rho)$
and a requested rate $R>r$. For all sufficiently large $n$ set
\begin{equation}
 m_n=\lfloor Rn\rfloor,
 \qquad
 \widetilde m=\left\lceil\frac{m_n}{r}\right\rceil.
 \label{eq:Mn}
\end{equation}
Since $r<E_D(\rho)$ and $\lfloor r\widetilde m\rfloor\ge m_n$,
there are LOCC distillation channels on $\widetilde m$
source copies producing $m_n$ Bell pairs with fidelity
tending to one, after discarding any excess output pairs. 

Let $\tau$ be a separable bipartite state on the same
Hilbert space as $\rho$. 
Choose an integer $N\geq 2$ such that
\begin{equation}
    \frac{nN}{\widetilde m}\in\mathbb{N},
    \qquad
    \frac{\log_2(N+1)}{N}\leq\delta_n.
\end{equation}
Such a finite $N$ exists because
$\log_2(N+1)/N\to0$ along the multiples of
$\widetilde m/\gcd(n,\widetilde m)$.
By Proposition~\ref{prop:decoupled}, there exists a 
correlated catalytic LOCC protocol with input
$\rho^{\otimes n}$ and output marginal
\begin{equation}
    \Omega_S
    =
    \frac{n}{\widetilde m}\rho^{\otimes\widetilde m}
    +
    \left(1-\frac{n}{\widetilde m}\right)
    \tau^{\otimes\widetilde m},
\end{equation}
satisfying
\begin{equation}
    \Omega_C=\omega_C,
    \qquad
    I(S:C)_\Omega
    \leq\frac{\log_2(N+1)}{N}
    \leq\delta_n.
\end{equation}

Now we proceed to the second stage. We apply the ordinary $\widetilde m$-copy distillation channel only to the system output. Let us denote it by $\mathcal{D}_d.$ This does not change the catalyst marginal and cannot increase $I(S:C)$. Additionally, we have
\begin{equation}
\begin{split}
    &\mathrm{Tr}\left[ \Ph^{\otimes m_n}\mathcal{D}_d(\Omega_S)  \right]=\frac{n}{\widetilde m}\mathrm{Tr}\left[ \Ph^{\otimes m_n}\mathcal{D}_d(\rho^{\otimes \widetilde m}) \right]+\\
    &\left( 1-\frac{n}{\widetilde m} \right)\mathrm{Tr}\left[ \Ph^{\otimes m_n}\mathcal{D}_d(\tau^{\otimes \widetilde m}) \right]    
\end{split}
\end{equation}

Since $r<\ED(\rho)$, ordinary distillation achieves
\begin{equation}
 \mathrm{Tr}\left[
 \Ph^{\otimes m_n}
 \mathcal{D}_d(\rho^{\otimes \widetilde m})
 \right]=1-o(1).
\end{equation}
Moreover, since $\tau$ is separable and $\mathcal{D}_d$
is LOCC, the state
$\mathcal{D}_d(\tau^{\otimes \widetilde m})$ is separable.
Thus, by Lemma~\ref{lem:sepoverlap},
\begin{equation}
 0\le
 \mathrm{Tr}\left[
 \Ph^{\otimes m_n}
 \mathcal{D}_d(\tau^{\otimes \widetilde m})
 \right]
 \le 2^{-m_n}=o(1).
\end{equation}
Consequently, the achieved fidelity satisfies
\begin{equation}
 \begin{aligned}
 \mathrm{Tr}\left[
 \Ph^{\otimes m_n}\mathcal{D}_d(\Omega_S)
 \right]
 &=\frac{n}{\widetilde m}(1-o(1))+o(1)\\
 &\longrightarrow\frac{r}{R},
 \end{aligned}
 \label{eq:rlower}
\end{equation}
where we used Eq. \eqref{eq:Mn}.

For $R>\ED(\rho)$, this construction therefore gives
\begin{equation}
 \liminf_{n\to\infty}
 \Fcc_{\rho\to\Ph}(n,m_n;\delta_n)
 \ge\frac{r}{R}.
\end{equation}
Since this holds for every fixed
$0<r<\ED(\rho)$, letting
$r\uparrow\ED(\rho)$ yields
\begin{equation}
 \liminf_{n\to\infty}
 \Fcc_{\rho\to\Ph}(n,m_n;\delta_n)
 \ge\frac{\ED(\rho)}{R}.
\end{equation}

For $R<\ED(\rho)$, ordinary unassisted
distillation already achieves fidelity tending to one.
At $R=\ED(\rho)$, the same catalytic construction
gives an asymptotic fidelity of at least
$r/\ED(\rho)$ for every fixed $r<\ED(\rho)$.
Letting $r\uparrow\ED(\rho)$ therefore yields
a lower bound of one.

Combining all three cases, we obtain
\begin{equation}
 \liminf_{n\to\infty}
 \Fcc_{\rho\to\Ph}(n,m_n;\delta_n)
 \ge
 \min\left\{1,\frac{\ED(\rho)}{R}\right\}.
 \label{eq:mixedlower}
\end{equation}

For every $R<\ED (\rho)$, the limiting fidelity is one; for every $R>\ED (\rho)$, it is $\ED (\rho)/R<1$.  Hence the reliable rate is $\ED (\rho)$.  For every finite $R$, the limiting fidelity is positive because $\ED (\rho)>0$. Therefore the nonvanishing-fidelity threshold is infinite, proving \eqref{eq:mixedrates}.
 
\section{Proof of Theorem \ref{thm::exp}}{\label{se::exp}}

In this section, we detail the proof of theorem~\ref{thm::exp}.
Let $f_n$ denote the fidelity of the catalytic protocol with the target state $\Phi_{2^{m_n}}.$ By assumption, we have $f_n>f_0.$ Since $m_n/n\to R>r_{\text{sc}}$, for all sufficiently large $n$,
we have $n\ge N_0$ and $m_n\ge r_{\text{sc}}n$.
Lemma~\ref{lem:auxiliary-emax} and
Eq.~\eqref{eq::strongexpconverse}
therefore give
\begin{equation}
    f_0
    \le f_n
    \le
    2^{E_{\max}(\omega_C)}
    F_{\rho\to \Ph}^{\text{not}}(n,m_n)
    \le
    2^{E_{\max}(\omega_C)-\gamma n}.
\end{equation}
Taking logarithms proves
Eq.~\eqref{eq:emax-linear}.

We now repeat the protocol \(k\) times, each time using a fresh input block \(\rho^{\otimes n}\) and the same catalyst. Since the catalyst marginal is returned exactly after each use, the resulting process is a catalytic LOCC protocol with input \(\rho^{\otimes kn}\), target \(km_n\) Bell pairs, and catalyst \(\omega_C\).
Since
\begin{equation}
    km_n\ge r_{\text{sc}}kn,
\end{equation}
our exponential-converse assumption implies
\begin{equation}
    F_{\rho\to \Ph}^{\text{not}}(kn,km_n)\le 2^{-\gamma kn}.
\end{equation}
Now, we impose the additional constraints that the correlations within the joint output state $\Omega_{SC}$ are restricted by
\begin{equation}
    I(S:C)_{\Omega}\le\delta_n,
    \quad
    0<\delta_n\longrightarrow0.
\end{equation}
Then, by applying Lemma~\ref{lem:reuse-emax}, we obtain
\begin{equation}
\label{eq:reuse-exponential}
    f_0^k
    \le f_n^k
    \le
    2^{E_{\max}(\omega_C)-\gamma kn}
    +
    \sqrt{\frac{\ln2}{2}(k-1)\delta_n}.
\end{equation}
Set
\begin{equation}
    k_n=
    \left\lfloor\frac{\ell_n}{4b}\right\rfloor\!,\,\,
    b=\log_2(1/f_0)>0,
    \,\,
    \ell_n=\log_2(1/\delta_n).
\end{equation}
Since $\delta_n\to0$, eventually $k_n\ge2$.
Moreover,
\begin{equation}
    f_0^{k_n}
    =
    2^{-bk_n}
    \ge
    2^{-\ell_n/4}
    =
    \delta_n^{1/4}.
\end{equation}
Consequently,
\begin{multline}
    \frac{
        \sqrt{\frac{\ln2}{2}(k_n-1)\delta_n}
    }{f_0^{k_n}}
    \le
    \sqrt{\frac{\ln2}{2}k_n}\,
    \delta_n^{1/4}\\ 
    \le \sqrt{\frac{\ln 2}{8b}\ell_n}2^{-\ell_n/4}
    \longrightarrow0,
\end{multline}
where we used the fact that exponential decay dominates polynomial growth.
Thus, for all sufficiently large $n$,
\begin{equation}
    \sqrt{\frac{\ln2}{2}(k_n-1)\delta_n}
    \le
    \frac12 f_0^{k_n}.
\end{equation}
Substituting $k=k_n$ into
Eq.~\eqref{eq:reuse-exponential} gives
\begin{equation}
    2^{E_{\max}(\omega_C)-\gamma k_n n}
    \ge
    \frac12 f_0^{k_n}.
\end{equation}
Taking logarithms yields
\begin{equation}
\label{eq:emax-explicit-cost}
    E_{\max}(\omega_C)
    \ge
    k_n(\gamma n-b)-1.
\end{equation}
For sufficiently large $n,$ $k_n\geq \frac{\ell_n}{8b}$ and $n\geq \frac{2b}{\gamma}.$ Consequently,
\begin{equation}
   E_{\max}(\omega_C) \geq \frac{\gamma}{16b}n\ell_n-1.
\end{equation}
The threshold in $n$ and the constants in this bound
depend only on the fixed parameters and the prescribed
sequences $(m_n)$ and $(\delta_n)$, not on the protocol
or catalyst. Thus, the bound holds uniformly over all
admissible protocols. Eq.~\eqref{eq:emax-correlation-cost} follows.

\section{Proof of Theorem~\ref{thm::achieveexp}}{\label{se::achieveexp}}
In this section, we detail the proof of theorem~\ref{thm::achieveexp}. As in the proof of Theorem~\ref{thm:mixedmain}, the protocol consists of
two stages. First, correlated catalysis transforms
$n$ copies of $\rho$ into a mixture of
$\rho^{\otimes\widetilde m}$, with $\widetilde m>n$,
and a separable state. Second, ordinary LOCC distillation
is applied to this output to produce $m_n$ approximate
Bell pairs. To reduce the catalyst memory cost, the first
stage uses a truncated binomial distribution instead of
the fixed-weight distribution used in Proposition~\ref{prop:decoupled}.

Choose fixed constants $p,r$ such that
\begin{equation}
    f_0<p<\frac{r}{R}<\frac{E_D(\rho)}{R}.
\end{equation}
For all sufficiently large $n$, set
\begin{equation}
    m_n=\lfloor Rn\rfloor,
    \qquad
    \widetilde m=
    \left\lceil\frac{m_n}{r}\right\rceil>n.
\end{equation}
Then
\begin{equation}
    \frac{n}{\widetilde m}\longrightarrow\frac{r}{R}>p.
\end{equation}
Since $r<E_D(\rho)$ and
$\lfloor r\widetilde m\rfloor\geq m_n$, there exist
deterministic LOCC distillation channels $\mathcal D_d$
such that
\begin{equation}
    \operatorname{Tr}\!\left[
        \Ph^{\otimes m_n}
        \mathcal D_d(\rho^{\otimes\widetilde m})
    \right]
    =1-\varepsilon_n,
    \qquad \varepsilon_n\to0.
\end{equation}
We now construct the first stage of the protocol.

Let $\tau$ be a separable state on the same bipartite
Hilbert space as $\rho$, and define
\begin{equation}
    a
    =
    \rho^{\otimes n}
    \otimes
    \tau^{\otimes(\widetilde m-n)}.
\end{equation}
The additional copies of $\tau$ can be prepared
freely by LOCC.

For an integer $N\ge2$, to be chosen below, set
\begin{equation}
    K=\left\lfloor\frac{nN}{\widetilde m}\right\rfloor.
\end{equation}
Let $Q_p$ be the probability distribution on
$\{0,1\}^N$ given by
\begin{equation}
    Q_p(x)=p^{|x|}(1-p)^{N-|x|},
\end{equation}
where $|x|=\sum_{i=1}^{N}x_i$ is the Hamming weight
of the binary string $x=(x_1,\ldots,x_N)$, i.e.,
the number of entries equal to one.
Define
\begin{equation}
    Z=\sum_{|x|\le K}Q_p(x),
    \qquad
    P(x)=
    \frac{Q_p(x)\mathbf 1_{\{|x|\le K\}}}{Z},
\end{equation}
where 
\begin{equation}
    \mathbf 1_{\{|x|\le K\}}
    :=
    \begin{cases}
        1, & |x|\le K,\\
        0, & |x|>K.
    \end{cases}
\end{equation}
Thus, \(P(x)\) is the probability of sampling the string \(x\).
Consider the permutation-invariant state
\begin{equation}
    \Gamma_N
    =
    \sum_{x\in\{0,1\}^N}
    P(x)\,
    \theta_{x_1}\otimes\cdots\otimes\theta_{x_N},
    \,\,
    \theta_1=\rho^{\otimes \widetilde m},\,\,\theta_0=\tau^{\otimes \widetilde m}.
\end{equation}
The transformation $a^{\otimes N}\to\Gamma_N$
is deterministic LOCC.
Indeed, Alice samples $x$ according to $P$ and
communicates it to Bob.
Every string in the support of $P$ requires
$|x|\widetilde m\le nN$ copies of $\rho$.
The parties discard the unused copies of $\rho$,
replace them with copies of $\tau$, and perform
matching local permutations to prepare the
corresponding tensor product.

Let $\Gamma_j$ denote the $j$-block marginal of
$\Gamma_N$, with $\Gamma_0=1$ (it is a scalar, see the comment in Lemma~\ref{lem:batch}) on the trivial system.
By Lemma~\ref{lem:batch}, the state
\begin{equation}
    \omega_C
    =
    \frac1N\sum_{j=0}^{N-1}
    |j\rangle\langle j|_J
    \otimes a^{\otimes j}\otimes\Gamma_{N-1-j}
\end{equation}
enables a correlated catalytic conversion
of one fresh block $a$ into an output with marginal
$\Gamma_1$.
The complete catalyst marginal is returned exactly,
and the output--catalyst mutual information at this
stage equals
\begin{equation}
   I(S:C)= \frac1N D(\Gamma_N\|\Gamma_1^{\otimes N}).
\end{equation}

To bound this quantity, define
\begin{equation}
    \sigma_p=p\rho^{\otimes \widetilde m}+(1-p)\tau^{\otimes \widetilde m}.
\end{equation}
Let $q_p(0)=1-p$ and $q_p(1)=p$, so that
$Q_p(x)=\prod_{i=1}^N q_p(x_i)$.
Denote by $\mathcal E$ the classical-to-quantum
channel that prepares
$\theta_{x_1}\otimes\cdots\otimes\theta_{x_N}$
conditional on the classical string $x$.
It maps the distribution $P$ to $\Gamma_N$
and $Q_p$ to $\sigma_p^{\otimes N}$.
Indeed,
\begin{align}
    \mathcal E(Q_p)
    &=
    \sum_{x_1,\ldots,x_N\in\{0,1\}}
    \left(\prod_{i=1}^N q_p(x_i)\right)
    \theta_{x_1}\otimes\cdots\otimes\theta_{x_N}
    \nonumber\\
    &=
    \bigotimes_{i=1}^N
    \left(\sum_{x_i=0}^1 q_p(x_i)\theta_{x_i}\right)
    =
    \bigl((1-p)\theta_0+p\theta_1\bigr)^{\otimes N}
    \nonumber\\
    &=\sigma_p^{\otimes N}.
\end{align}
Data processing therefore gives
\begin{equation}
    D(\Gamma_N\|\sigma_p^{\otimes N})
    \le
    D(P\|Q_p)
    =
    -\log_2 Z.
\end{equation}
Furthermore, since every one-block marginal
of $\Gamma_N$ equals $\Gamma_1$,
\begin{align}
    D(\Gamma_N\|\sigma_p^{\otimes N})
    &=
    D(\Gamma_N\|\Gamma_1^{\otimes N})
    +N D(\Gamma_1\|\sigma_p).
\end{align}
Consequently,
\begin{equation}
    \frac1N D(\Gamma_N\|\Gamma_1^{\otimes N})
    \le
    -\frac1N\log_2 Z.
    \label{eq:truncated-batch-correlation}
\end{equation}

Choose a fixed $\eta>0$ such that, for all
sufficiently large $n$,
\begin{equation}
    \frac n{\widetilde m}\ge p+\eta.
\end{equation}
For $X\sim\operatorname{Bin}(N,p)$, Hoeffding's
inequality yields~\cite{Hoeffdinginequality}
\begin{align}
    1-Z
    &=
    \Pr(X>K)
    \nonumber\\
    &=
    \Pr\!\left(\frac XN>\frac n{\widetilde m}\right)
    \nonumber\\
    &\le e^{-2\eta^2N}.
\end{align}
Whenever $1-Z\le1/2$, we also have
\begin{equation}
    -\log_2 Z
    \le
    \frac{2}{\ln2}(1-Z).
\end{equation}
Combining these inequalities with
Eq.~\eqref{eq:truncated-batch-correlation} gives
\begin{equation}
    \frac1N D(\Gamma_N\|\Gamma_1^{\otimes N})
    \le
    \frac{2}{N\ln2}e^{-2\eta^2N}.
    \label{eq:exponential-batch-correlation}
\end{equation}

The condition $Z\ge1/2$ means that the admissible
configurations carry at least half of the probability
under the original distribution $Q_p$.
Since $1-Z\le e^{-2\eta^2N}$, it is sufficient to take
\begin{equation}
    N\ge\frac{\ln2}{2\eta^2}.
\end{equation}
As $\eta>0$ is fixed, this condition is automatically
satisfied for sufficiently large $N$.
It imposes no additional assumption on the source state.
Note that $Z$ is a normalization factor, not the success
probability of the catalytic protocol: the protocol
samples directly from $P$, whose support contains
only admissible configurations.

For all sufficiently large $n$, we have $\delta_n<1$.
We may then choose
\begin{equation}
    N
    =
    \max\left\{
        2,\
        \left\lceil
        \frac{1}{2\eta^2}
        \ln\!\left(\frac{2}{\delta_n\ln2}\right)
        \right\rceil
    \right\}.
\end{equation}
Here $\ln$ denotes the natural logarithm.
This choice ensures
\begin{equation}
    \frac1N D(\Gamma_N\|\Gamma_1^{\otimes N})
    \le\delta_n,
    \qquad
    N=O\!\left(\log_2\frac1{\delta_n}\right),
\end{equation}
and $N\to\infty$.

By permutation invariance,
\begin{equation}
    \Gamma_1=p_N\rho^{\otimes \widetilde m}+(1-p_N)\tau^{\otimes \widetilde m},
    \qquad
    p_N=\frac1N\mathbb E[X\mid X\le K].
\end{equation}
Since conditioning on $X\le K$ removes only an
upper tail,
\begin{equation}
    0\le p-p_N\le1-Z\le e^{-2\eta^2N}.
\end{equation}
Thus $p_N\to p$.

Finally, apply $\mathcal{D}_d$ (recall that $\mathcal{D}_d$ is ordinary distillation protocol) only to the released
output block.
This preserves the catalyst marginal and cannot
increase the output--catalyst mutual information.
The final state therefore satisfies
\begin{equation}
    \Omega_C=\omega_C,
    \qquad
    I(S:C)_\Omega\le\delta_n,
    \qquad
    \Omega_S=\mathcal D_d(\Gamma_1).
\end{equation}
Because $\mathcal{D}_d(\tau^{\otimes \widetilde m})$ is separable,
\begin{equation}
    0\le
    \operatorname{Tr}\!\left[
        \Phi_{2^{m_n}}\mathcal{D}_d(\tau^{\otimes \widetilde m})
    \right]
    \le2^{-m_n}.
\end{equation}
Hence
\begin{equation}
    p_N(1-\varepsilon_n)
    \le f_n
    \le p_N+(1-p_N)2^{-m_n},
\end{equation}
which implies
\begin{equation}
    f_n\longrightarrow p>f_0.
\end{equation}
In particular, $f_n\ge f_0$ for all sufficiently
large $n$.
No comparison between $\varepsilon_n$ and
$\delta_n$ is required.

It remains to bound the catalyst dimensions.
Let $d_A,d_B$ be the local dimensions of $\rho$.
Each catalyst block contains $\widetilde m$
copies of the source Hilbert space, and the
classical clock has $N$ values.
Including the clock, the local dimensions can
be chosen to satisfy
\begin{align}
    \log_2\dim C_A
    &\le
    \widetilde m(N-1)\log_2 d_A+\log_2N,
    \\
    \log_2\dim C_B
    &\le
    \widetilde m(N-1)\log_2 d_B+\log_2N.
\end{align}
Since $\widetilde m=O(n)$ and
$N=O(\log_2(1/\delta_n))$, both quantities are
$O(n\log_2(1/\delta_n))$, as claimed.

\section{AUXILIARY LEMMAS AND PROPOSITIONS}{\label{se::theorem2lemmas}}
\begin{lemma}[Separable target overlap~\cite{Khatripqct}]
\label{lem:sepoverlap}
For every separable state $\sigma_{AB}$ with local dimension $D$,
\begin{equation}
 \Tr(\Phi_D\sigma)\le\frac1D.
 \label{eq:sepoverlap}
\end{equation}
\end{lemma}

\begin{lemma}[Distillable entanglement of a finite catalytic output]
\label{lem:EDoutput}
Assume that $E_D(\rho)>0.$ Suppose a finite-block exact-return catalytic LOCC protocol maps $\rho^{\otimes n}$ to an output marginal $\Omega_S$.  Then
\begin{equation}
 \ED(\Omega_S)\le n\ED(\rho).
 \label{eq:EDoutput}
\end{equation}
\end{lemma}

\begin{proof}
Fix $n$ and initial catalyst $\omega_C$ and assume instead that $\ED(\Omega_S)>nE_D(\rho)$. Consequently, there exists $r$ such that $n \ED(\rho)<r<\ED (\Omega_S).$ Run $t$ independent copies of the finite protocol on $\rho^{\otimes n}$, using independent copies of its finite catalyst. The system output is exactly $\Omega_S^{\otimes t}$ and the complete catalyst marginal is exactly $\omega_C^{\otimes t}$.  Apply an ordinary distillation protocol only to $\Omega_S^{\otimes t}$. Such protocol does not change the catalyst marginal. Since $r<\ED (\Omega_S)$ there exist protocol which distill $\lfloor rt\rfloor$ copies of $\Ph$ state as $t\to\infty.$ Thus, we constructed a protocol which consumes $nt$ copies of a state $\rho$ and distills $rt$ copies of $\Ph.$ The rate of this protocol is $\frac{\lfloor rt\rfloor}{nt}=\frac{r}{n}>\ED (\rho)$ in the limit $t\rightarrow\infty.$ This contradicts the catalytic rate-invariance theorem
(Theorem~2 of Ref.~\cite{Ganardicatalysis}). Therefore, \eqref{eq:EDoutput} holds. 
\end{proof}

\begin{lemma}[Bell-product overlap lower-bounds $\ED$]
\label{lem:filter}
Let $\sigma$ be a state on $m\ge8$ qubit pairs and let
\begin{equation}
 f=\Tr(\Phi_{2^m}\sigma).
\end{equation}
Set
\begin{equation}
 t=\lceil2\log m\rceil,\qquad \ell=m-t>0.
 \label{eq::tde}
\end{equation}
Then
\begin{equation}
 \ED(\sigma)\ge f\ell-2
 =f\bigl(m-\lceil2\log m\rceil\bigr)-2.
 \label{eq:filterbound}
\end{equation}
\end{lemma}

\begin{proof}
Let $D=2^m$.  Apply the isotropic twirl $U\otimes\overline U$, which is LOCC and preserves $f$.  Since distillable entanglement cannot increase under LOCC, it is enough to lower-bound the distillable entanglement of
\begin{equation}
 \sigma_f=f\Phi_D+u(I-\Phi_D),
 \qquad u=\frac{1-f}{D^2-1},
 \label{eq:isotropic}
\end{equation}
Decompose each local $D$-dimensional space into two subsystems of dimensions $d_1$ and $d_2$ such that:
\begin{equation}
 d_1=2^t,\qquad d_2=2^\ell,\qquad D=d_1d_2.
\end{equation}
The maximally entangled state factors as $\Phi_D=\Phi_{d_1}\otimes\Phi_{d_2}$.

Alice and Bob measure their $d_1$-dimensional subsystems
in the computational basis. They keep the
$d_2$-dimensional subsystems exactly when their outcomes
agree and forget the common outcome. The ideal component $\Phi_D$ always passes and leaves $\Phi_{d_2}$.  The identity operator has $d_1$ matching outcomes, so the unnormalized retained state is
\begin{equation}
 X=f\Phi_{d_2}+u(d_1I_{d_2^2}-\Phi_{d_2}).
 \label{eq:filterX}
\end{equation}
Its success probability and unnormalized Bell overlap are
\begin{align}
 s&=\Tr X=f+u(d_1d_2^2-1),\label{eq:sprob}\\
 h&=\Tr(\Phi_{d_2}X)=f+u(d_1-1).\label{eq:hnum}
\end{align}
The normalized successful state is isotropic with local dimension $d_2$, with Bell fidelity
\begin{equation}
 \alpha=\frac hs.
\end{equation}
Its local marginals are maximally mixed, and therefore its coherent information is 
\begin{align}
 I(A\rangle B)_{X/s}
 &=\ell-h_2(\alpha)-(1-\alpha)\log(d_2^2-1)\\
 &\ge \ell-1-2\ell(1-\alpha)\\
 &=(2\alpha-1)\ell-1,
 \label{eq:coherentlower}
\end{align}
where $h_2$ denotes the binary entropy.
Here, $A$ and $B$ denote Alice's and Bob's retained
subsystems, respectively, each of dimension $d_2$.
The hashing inequality states that the positive part of coherent information is an achievable one-way distillation rate~\cite{Devetak2005}.  Applying the filter independently to many copies and hashing the successful outputs gives
\begin{align}
 &\ED(\sigma)\ge \ED(\sigma_f)\nonumber\\
 &\ge s\max\{0,(2\alpha-1)\ell-1\}\ge (2h-s)\ell-s.
 \label{eq:hashstep}
\end{align}
The final inequality is valid even when the expression is negative, because $\max\{0,x\}\ge x$.

We now estimate the last expression.  First, $h\ge f$, see Eq. \eqref{eq:hnum}.  Second,
\begin{equation}
 u(d_1d_2^2-1)
 =\frac{(1-f)(d_1d_2^2-1)}{d_1^2d_2^2-1}
 \le\frac{1}{d_1}=2^{-t}.
\end{equation}
Plugging it into Eq. \eqref{eq:sprob}, we obtain that $s\le f+2^{-t}$. Since also $s\le1$,
\begin{align}
 (2h-s)\ell-s
 &\ge(f-2^{-t})\ell-1\\
 &\ge f\ell-2,
\end{align}
where the second inequality uses the fact that $2^{-t}\ell\le m^{-2}m\le1$. Recalling \eqref{eq::tde} completes the proof of \eqref{eq:filterbound}.
\end{proof}

The following construction is a variant of the catalysts introduced in~\cite{Duancatalysis,Shiraishicorrelated}.
\begin{lemma}[Batch catalyst and entropy identity]
\label{lem:batch}
Let $a$ be an arbitrary bipartite state, and
suppose that a deterministic LOCC channel
transforms $a^{\otimes N}$ exactly into a
permutation-invariant $N$-block state $\Gamma_N$.
Denote by $\Gamma_j$ its $j$-block marginal
and set $\Gamma_0=1$ (it is a scalar, since all the subsystems are traced out).

Define a catalyst of $N-1$ blocks and a
classical clock by
\begin{equation}
\omega_C=\frac1N\sum_{j=0}^{N-1}
|j\rangle\langle j|_J\otimes
a^{\otimes j}\otimes\Gamma_{N-1-j}.
\label{eq:batchcat}
\end{equation}
Then, there is an exact catalytic conversion of one fresh block $a$ to an output with marginal $\Gamma_1$.  Moreover,
\begin{equation}
 I(S:C)_\Omega
 =S(\Gamma_1)-\frac1N S(\Gamma_N)
 =\frac1N D(\Gamma_N\Vert\Gamma_1^{\otimes N}),
 \label{eq:batchidentity}
\end{equation}
where $D(\rho\Vert \sigma)$ is quantum relative entropy.
\end{lemma}

\begin{proof}
We construct the LOCC protocol that implements the claimed transformation. Apply the following channel. For clock value $j<N-1$, regard one block of the $\Gamma_{N-1-j}$ part as the output, place the fresh $a$ into the bank, and increment the clock. The catalyst marginal in the new clock branch is
\begin{equation}
 a^{\otimes(j+1)}\otimes\Gamma_{N-2-j},
\end{equation}
which is precisely the term of \eqref{eq:batchcat} corresponding to $j+1$. Correlations between the released block and the retained $\Gamma$ blocks are kept; only the marginal of the retained blocks is $\Gamma_{N-2-j}$.

For clock value $j=N-1$, the bank and fresh input contain $a^{\otimes N}$.  Apply the assumed exact channel to obtain $\Gamma_N$, release one block, retain the remaining $N-1$ blocks, and reset the clock to zero.  The retained marginal is $\Gamma_{N-1}$, again the required next catalyst branch.  Since the clock is uniformly distributed and shifted cyclically, the complete catalyst marginal returns exactly.  Every released block has marginal $\Gamma_1$ by permutation invariance.

It remains to compute the correlation.  Put $S(\Gamma_0)=0.$
Orthogonality of the classical clock labels and standard properties of von Neumann entropy give:
\begin{align}
 S(\omega_C)
 &=\log N+\frac1N\sum_{j=0}^{N-1}\big(jS(a)+S(\Gamma_{N-1-j})\big),\label{eq:catentropy}\\
 S(\Omega_{SC})
 &=\log N+\frac1N\sum_{j=0}^{N-1}\big(jS(a)+S(\Gamma_{N-j})\big).
 \label{eq:jointentropy}
\end{align}
The second expression is indexed by the final clock value; at clock zero its conditional joint state is $\Gamma_N$, and at clock $j>0$ it is $a^{\otimes j}\otimes\Gamma_{N-j}$ with one $\Gamma$ block designated as $S$.

Since the catalyst returns exactly and $\Omega_S=\Gamma_1$,
\begin{align}
 I(S:C)_\Omega
 &=S(\Gamma_1)+S(\omega_C)-S(\Omega_{SC})\nonumber\\
 &=S(\Gamma_1)-\frac{S(\Gamma_N)}{N},
 \label{eq::isclem13}
\end{align}
where the sums telescope.  Finally, all one-block marginals of $\Gamma_N$ are $\Gamma_1$, so
\begin{equation}
\begin{split}
 &D(\Gamma_N\Vert\Gamma_1^{\otimes N})
 =\\
 &-S(\Gamma_N)-\Tr\Gamma_N\log(\Gamma_1^{\otimes N})
 =NS(\Gamma_1)-S(\Gamma_N).   
\end{split}
\end{equation}
Dividing by $N$ and comparing with Eq. \eqref{eq::isclem13} proves \eqref{eq:batchidentity}.
\end{proof}

\begin{proposition}[Decoupled source-state amplifier]
\label{prop:decoupled}
Let $\rho$ be a finite-dimensional bipartite state and
let $\tau$ be a separable state on the same Hilbert space.
Let $n,\widetilde m,N$ be positive integers satisfying
$\widetilde m>n$, $N\geq2$, and
$K=nN/\widetilde m\in\mathbb{N}$.
There exists a deterministic correlated catalytic
LOCC protocol with input $\rho^{\otimes n}$ and output
marginal
\begin{equation}
    \Omega_S
    =
    \frac{n}{\widetilde m}\rho^{\otimes\widetilde m}
    +
    \left(1-\frac{n}{\widetilde m}\right)
    \tau^{\otimes\widetilde m}.
    \label{eq:amplifier_output}
\end{equation}
The catalyst is returned exactly, $\Omega_C=\omega_C$,
and the residual correlations satisfy
\begin{equation}
\begin{aligned}
    I(S:C)_\Omega
    &\leq h_2\!\left(\frac{n}{\widetilde m}\right)
    -\frac{1}{N}\log_2
      \binom{N}{K}\\
    &\leq \frac{\log_2(N+1)}{N}.
\end{aligned}
    \label{eq:amplifier_correlations}
\end{equation}
\end{proposition}

\begin{proof}
Set
\begin{equation}
    a_1=\rho^{\otimes n}
        \otimes\tau^{\otimes(\widetilde m-n)},
    \qquad
    K=\frac{nN}{\widetilde m}.
\end{equation}
Since $\tau$ is separable, the additional copies of
$\tau$ can be prepared freely by LOCC.
Define
\begin{equation}
    \theta_1=\rho^{\otimes\widetilde m},
    \qquad
    \theta_0=\tau^{\otimes\widetilde m},
\end{equation}
and the permutation-invariant state
\begin{equation}
    \Gamma_N
    =
    \frac{1}{\binom{N}{K}}
    \sum_{\substack{x\in\{0,1\}^N\\ |x|=K}}
    \theta_{x_1}\otimes\cdots\otimes\theta_{x_N},
    \label{eq:amplifier_batch}
\end{equation}
where $|x|=\sum_{i=1}^N x_i$.
The state $a_1^{\otimes N}$ contains exactly $nN$
copies of $\rho$ and $(\widetilde m-n)N$ copies of $\tau$.
Every tensor product term in
Eq.~\eqref{eq:amplifier_batch} contains exactly
$K\widetilde m=nN$ copies of $\rho$ and
$(N-K)\widetilde m=(\widetilde m-n)N$ copies of $\tau$.

Alice chooses a binary string of weight $K$
uniformly at random and communicates it to Bob.
Both parties then perform matching local
permutations to arrange the copies of $\rho$
and $\tau$ into the corresponding tensor product
term in Eq.~\eqref{eq:amplifier_batch}.
Consequently, $a_1^{\otimes N}\to\Gamma_N$
is a deterministic LOCC transformation.

Since $\Gamma_N$ is permutation invariant,
every one-block marginal is identical.
The probability that a given block contains
$\rho^{\otimes \widetilde m}$ equals $K/N=n/\widetilde m$.
Therefore,
\begin{equation}
 \Gamma_1=
 \frac {n}{\widetilde m} \rho^{\otimes \widetilde m}+\left(1-\frac {n}{\widetilde m}\right)\tau^{\otimes \widetilde m}.
\end{equation}
By Lemma~\ref{lem:batch}, the transformation
$a_1^{\otimes N}\to\Gamma_N$ yields a correlated
catalytic conversion of $a_1$ into an output with
marginal $\Omega_S=\Gamma_1$.
Since the additional copies of $\tau$ can be prepared
freely, this gives the claimed protocol with input
$\rho^{\otimes n}$ and proves Eq.~\eqref{eq:amplifier_output}.

By Lemma~\ref{lem:batch}, the residual
correlation is
\begin{equation}
 I(S:C)_\Omega
 =\frac1N
 D(\Gamma_N\Vert\Gamma_1^{\otimes N}).
 \label{eq:batchrelative}
\end{equation}
Let $p=K/N=n/\widetilde m$, and let $P_{N,K}$ denote
the uniform probability distribution over
binary strings of length $N$ and weight $K$.
Let $Q_p=\mathrm{Bern}(p)^{\otimes N}$
be the distribution of $N$ independent
Bernoulli random variables.
Consider the classical-to-quantum channel
$\mathcal{E}$ that prepares
$\theta_{x_1}\otimes\cdots\otimes\theta_{x_N}$
conditioned on a classical string
$x=(x_1,\ldots,x_N)$.
By construction,
\begin{equation}
 \mathcal{E}(P_{N,K})=\Gamma_N,
 \qquad
 \mathcal{E}(Q_p)=\Gamma_1^{\otimes N}.
\end{equation}

The data-processing inequality for quantum
relative entropy therefore gives
\begin{align}
 D(\Gamma_N\Vert\Gamma_1^{\otimes N})
 &\le D(P_{N,K}\Vert Q_p) \nonumber\\
 &=-\log_2\left[
 \binom NK p^K(1-p)^{N-K}
 \right] \nonumber\\
 \label{eq:batchDPI}
\end{align}

Finally, define
\begin{equation}
 q_K=\binom NKp^K(1-p)^{N-K}.
\end{equation}
Since $K=Np$ is a mode of the binomial
distribution, and the distribution has
$N+1$ possible outcomes,
\begin{equation}
 q_K\ge\frac1{N+1}.
\end{equation}
Consequently, by Eq. \eqref{eq:batchDPI}
\begin{equation}
 I(S:C)_\Omega
 \le-\frac1N\log_2 q_K
 \le\frac{\log_2(N+1)}N,
\end{equation}
which proves Eq.~\eqref{eq:amplifier_correlations}.
\end{proof}
\begin{lemma}[Auxiliary-state bound]
\label{lem:auxiliary-emax}
Let $\Lambda$ be a deterministic LOCC channel acting on
$\rho^{\otimes n}\otimes\omega_C$, where $\omega_C$ is a
finite-dimensional bipartite auxiliary state. Write
\begin{equation}
    \Omega_{SC}
    =
    \Lambda(\rho^{\otimes n}\otimes\omega_C),
    \qquad
    f=\operatorname{Tr}(\Phi_{2^{m}}\Omega_S).
    \label{eq::fomega}
\end{equation}
Then
\begin{equation}
\label{eq:auxiliary-emax}
    f
    \le
    \min\!\left\{
        1,\,
        2^{E_{\max}(\omega_C)}
        F_{\rho\to\Ph}^{\text{not}}(n,m)
    \right\}.
\end{equation}
No return condition on the auxiliary state is required.
\end{lemma}

\begin{proof}
Fix $\eta>0$. By the definition of $E_{\max}$ (see \eqref{eq:Emaxdef}), there
exists a normalized separable state $\tau_C$ such that~\cite{Datteentropies}
\begin{equation}
    \omega_C
    \le
    2^{E_{\max}(\omega_C)+\eta}\tau_C.
\end{equation}
Tensoring with $\rho^{\otimes n}$ and applying the
positive map $\Lambda$ gives
\begin{equation}
    \Lambda(\rho^{\otimes n}\otimes\omega_C)
    \le
    2^{E_{\max}(\omega_C)+\eta}
    \Lambda(\rho^{\otimes n}\otimes\tau_C).
\end{equation}
Consequently,
\begin{align}
    f
    &\le
    2^{E_{\max}(\omega_C)+\eta}
    \operatorname{Tr}\!\left[
        (\Phi_{2^{m}}\otimes I_C)
        \Lambda(\rho^{\otimes n}\otimes\tau_C)
    \right]
    \nonumber\\
    &\le
    2^{E_{\max}(\omega_C)+\eta}
    F_{\rho\to\Ph}^{\text{not}}(n,m).
\end{align}
Indeed, preparing the separable state $\tau_C$, applying
$\Lambda$, and discarding $C$ defines a non-catalytic
LOCC protocol. Taking $\eta\downarrow0$ and using
$f\le1$ proves the claim.
\end{proof}

\begin{lemma}[Repeated-use bound]
\label{lem:reuse-emax}
Under the assumptions of Lemma~\ref{lem:auxiliary-emax},
suppose additionally that
\begin{equation}
    \Omega_C=\omega_C,
    \qquad
    I(S:C)_\Omega\le\delta.
\end{equation}
Then, for every integer $k\ge2$,
\begin{equation}
\label{eq:reuse-emax}
    f^k
    \le
    2^{E_{\max}(\omega_C)}
    F_{\rho\to\Ph}^{\text{not}}(kn,mk)
    +
    \sqrt{\frac{\ln 2}{2}(k-1)\delta}.
\end{equation}
\end{lemma}

\begin{proof}
Apply the same protocol $k$ times, using fresh input
blocks $\rho^{\otimes n}$ and the same catalyst,
without acting again on previous outputs.
Let $S_i$ denote the entire output block of use $i$,
including both parties' systems, and let
$\Theta_{S_1\cdots S_k}$ denote the final joint state
of all output blocks. Each use produces the same output marginal, so $\Theta_{S_i}=\Omega_S.$ Note that the total output might be correlated and in general $\Theta_{S_1\cdots,S_k}\neq \Omega_S^{\otimes k}.$

Before each use, the catalyst marginal is $\omega_C$,
and the fresh input is independent of the catalyst
and all previous outputs. The reduced input to each
use is therefore $\rho^{\otimes n}\otimes\omega_C$.
Consequently, immediately after each use, the joint
marginal of the new output and the catalyst is
\begin{equation}
    \Omega_{S_iC}
    =
    \Lambda(\rho^{\otimes n}\otimes\omega_C)
    =
    \Omega_{SC},
\end{equation}
up to relabeling $S$ as $S_i$. In particular,
\begin{equation}
    \Omega_C=\omega_C,
    \qquad
    I(S_i:C)_{\Omega}
    =
    I(S:C)_{\Omega}
    \le\delta.
\end{equation}
Here $\Omega_{S_iC}$ refers to the state immediately
after use $i$. Since subsequent uses do not act on
$S_i$, its final marginal also satisfies
$\Theta_{S_i}=\Omega_S$.

Fix $i<k$. Starting immediately after use $i$, all
future outputs are obtained by processing the catalyst
together with fresh independent input blocks.
Including these fixed input states in the channel
definition, the remaining protocol followed by
discarding the final catalyst is a quantum channel
\begin{equation}
    \mathcal{T}_i:C\longrightarrow S_{i+1}\cdots S_k.
\end{equation}
Thus
\begin{equation}
    \Theta_{S_i\cdots S_k}
    =
    (\operatorname{id}_{S_i}\otimes\mathcal{T}_i)
    (\Omega_{S_iC}).
\end{equation}
Data processing gives
\begin{equation}
    I(S_i:S_{i+1}\cdots S_k)_{\Theta}
    \le
    I(S_i:C)_{\Omega}
    \le\delta,
    \label{eq::idelta}
\end{equation}
where each mutual information is evaluated on the
corresponding joint state.

Using $\Theta_{S_i}=\Omega_S$ and the entropy chain
identity, we obtain (recall that $D$ is quantum relative entropy.)
\begin{align}
    D\!\left(
        \Theta_{S_1\cdots S_k}
        \middle\|
        \Omega_S^{\otimes k}
    \right)
    &=
    \sum_{i=1}^k S(\Theta_{S_i})
    -
    S(\Theta_{S_1\cdots S_k})
    \nonumber\\
    &=
    \sum_{i=1}^{k-1}
    I(S_i:S_{i+1}\cdots S_k)_{\Theta}
    \le(k-1)\delta.
\end{align}
In the last step we used Eq. \eqref{eq::idelta}.

Quantum Pinsker's inequality therefore implies~\cite{Hiaiinequality}
\begin{equation}
    \frac12
    \left\|\Theta-\Omega_S^{\otimes k}\right\|_1
    \le
    \sqrt{\frac{\ln2}{2}(k-1)\delta}.
    \label{eq::lem11Pin}
\end{equation}

 Eq. \eqref{eq::fomega} gives
$\mathrm{Tr}(\Phi_{2^{m}}^{\otimes k}\Omega_S^{\otimes k})=f^{k}.$ Since
$0\le\Phi_{2^{m}}^{\otimes k}\le I$, the trace-distance
bound yields
\begin{equation}
    \begin{split}
        &\vert f^{k} - \mathrm{Tr}(\Phi_{2^{m}}^{\otimes k}\Theta_{S_1\cdots S_k}) \vert= \\
        &\vert \mathrm{Tr}(\Phi_{2^{m}}^{\otimes k}\Omega_S^{\otimes k})-\mathrm{Tr}(\Phi_{2^{m}}^{\otimes k}\Theta_{S_1\cdots S_k})\vert \leq
        \frac{1}{2}\Vert \Omega_S^{\otimes k} - \Theta_{S_1\cdots S_k}  \Vert_1
    \end{split}
\end{equation}
Using the fact that $|a-b|\leq c\Rightarrow a\le b+c$ we obtain:
\begin{equation}
    f^{k}\leq \mathrm{Tr}(\Phi_{2^{m}}^{\otimes k}\Theta_{S_1\cdots S_k})+\frac{1}{2}\Vert \Omega_S^{\otimes k} - \Theta_{S_1\cdots S_k}  \Vert_1
    \label{eq::fkabc}
\end{equation}

The complete repeated-use process is a deterministic
LOCC protocol acting on
$\rho^{\otimes kn}\otimes\omega_C$, with the catalyst
supplied only once. Since $\Phi_{2^{m}}^{\otimes k}$ is
locally unitarily equivalent to $\Phi_{2^{mk}}$,
the auxiliary-state bound gives
\begin{equation}
    \operatorname{Tr}(\Phi_{2^{m}}^{\otimes k}\Theta)
    \le
    2^{E_{\max}(\omega_C)}
    F_{\rho\to\Ph}^{\text{not}}(kn,mk).
    \label{eq::lem11fi}
\end{equation}
Combining Eqs. \eqref{eq::lem11fi}, \eqref{eq::fkabc} and \eqref{eq::lem11Pin}, we obtain:
\begin{equation}
    f^k
    \le
    2^{E_{\max}(\omega_C)}
    F_{\rho\to\Ph}^{\text{not}}(kn,mk)
    +
    \sqrt{\frac{\ln2}{2}(k-1)\delta}.
\end{equation}
\end{proof}

\section{Conclusions}

In this work, we have shown that correlated catalysis removes the
strong converse for entanglement distillation from every
finite-dimensional distillable bipartite state.
At any target rate $R>0$, the optimal asymptotic 
fidelity is $\min\{1,E_D(\rho)/R\}$, even under an arbitrarily
small positive bound on the output--catalyst mutual
information tending to zero.
The catalyst is returned exactly for every input blocklength.
Consequently, catalysis preserves the reliable distillation
rate while allowing nonvanishing fidelity at every finite rate.

An analogous catalytic enhancement occurs in coherence
manipulation under time-translation covariant operations~\cite{Shiraishicatalysis}.
Under suitable conditions on the coherent modes,
correlated catalysts can increase the reliable distillation
rate from zero to infinity~\cite{Shiraishicatalysis}.
Our result identifies a different effect for entanglement:
for every distillable source, the reliable rate remains
$E_D(\rho)$, while the strong-converse threshold becomes
infinite, even with asymptotically vanishing
output--catalyst mutual information.

The construction used in the proof of Theorem~\ref{thm:mixedmain} establishes
the optimal asymptotic fidelity, but requires a larger
catalyst memory. This motivates our second construction,
which reduces the memory cost while retaining any fixed
fidelity requirement $0<f_0<E_D(\rho)/R$ at rates
$R>E_D(\rho)$. For every distillable source, it requires
only $O(n\log_2(1/\delta_n))$ qubits of local catalyst
memory while ensuring that the fidelity $f_n$ with the
target state satisfies $f_n\geq f_0$ and that 
$I(S:C)_\Omega\leq\delta_n\to0$.
Whenever non-catalytic distillation obeys an exponential
strong converse at a threshold below $R$, a matching
lower bound holds uniformly over all admissible protocols.
Under this assumption, the optimal local memory cost
for such fixed fidelity requirements is therefore
$\Theta(n\log_2(1/\delta_n))$.

The fixed fidelity requirement can be chosen arbitrarily
close to $E_D(\rho)/R$, but the constants in the memory
bound and the required input blocklength may depend on
this choice. Thus, although the construction of
Theorem~\ref{thm:mixedmain} establishes the optimal fidelity limit, our
matching memory bounds do not determine the optimal
memory cost when $f_n\to E_D(\rho)/R$. Characterizing
the joint dependence of this cost on the fidelity gap
and the allowed output--catalyst correlations remains
an open question. 

\section*{Acknowledgment}
This work was supported by the National Science Centre Poland (Grant No. 2022/46/E/ST2/00115 and 2024/55/B/ST2/01590).

ChatGPT 6 Astra was used during the preparation of this manuscript. The authors checked and verified all the results.

\bibliography{references}
\end{document}